\documentclass[twoside]{article}
\usepackage{fleqn,espcrc2}
\usepackage{graphics,epsfig}

\def\etal{{\it et~al.}}
\def\beq{\begin{equation}}
\def\enq{\end{equation}}
\def\bea{\begin{eqnarray}}
\def\ena{\end{eqnarray}}
\def\bec{\begin{center}}
\def\enc{\end{center}}

\def\Mesz{M\'esz\'aros~}
\def\Pacz{Paczy\'nski~}
\def\nsns{NS-NS~}
\def\bhns{BH-NS~}
\def\nonum{\nonumber}
\def\eps{\epsilon}
\def\varep{\varepsilon}

\def\Fnu{F_{\nu}}

\def\siml{\lower4pt \hbox{$\buildrel < \over \sim$}}
\def\simg{\lower4pt \hbox{$\buildrel > \over \sim$}}
\def\Omj{\Omega_j}
\def\msun{M_\odot}
\def\Mbh{M_{bh}}

\title{ Gamma-Ray Bursts and Bursters}

\author{ P. \Mesz\address{Dept. of Astronomy \& Astrophysics,
                           Pennsylvania State University,
                           University Park, PA 16802, USA}%
\thanks{Also Institute for Theoretical Physics, UCSB, and 
Ctr. for Gravitational Physics \& Geometry, Penn State Univ. 
Supported by NASA NAG-5 2857 and NSF PHY94-07194}
\thanks{Plenary talk given at the 19th Texas Symp. on Relativistic Astrophysics
\& Cosmology, Paris, Dec. 1998}
}

\begin{document}

\begin{abstract}
Major advances have been made in the field of gamma-ray bursts in the last
two years. The successful discovery of X-ray, optical and radio afterglows, which 
were predicted by theory, has made possible the identification of host galaxies at 
cosmological distances. The energy release inferred in these outbursts place 
them among the most energetic and violent events in the Universe.
Current models envisage this to be the outcome of a cataclysmic event 
leading to a relativistically expanding fireball, in which particles
are accelerated at shocks and produce nonthermal radiation.
The substantial agreement between observations and the theoretical predictions 
of the standard fireball shock model provide confirmation of the basic aspects 
of this scenario. The continued observations show a diversity of behavior, 
providing valuable constraints for more detailed, post-standard models which
incorporate more realistic physical features.
Crucial questions being now addressed are the beaming at different energies
and its implications for the energetics, the time structure of the afterglow, 
its dependence on the central engine or progenitor system behavior, and the role 
of the environment on the evolution of the afterglow.
\vspace{1pc}
\end{abstract}

\maketitle

\section{\bf INTRODUCTION}
\label{sec:intro}

The discovery of gamma-ray bursts (GRB) by the Vela military test-ban treaty 
satellites was announced in 1973 \cite{kle73}, and was quickly confirmed by
Soviet Konus satellite measurements \cite{maz74}. Then, for 23 years GRB remained 
essentially just that: brief outbursts of gamma-rays which pierced, for a brief 
instant, an otherwise pitch-black gamma-ray sky. An intense debate festered for a 
long time on whether they were objects in our galaxy or at cosmological distances.
The first major breakthrough came in 1992 with the launch of the Compton Gamma-Ray 
Observatory, whose superb results are summarized in a review by 
Fishman \& Meegan \cite{fm95}. In particular the all-sky survey from the BATSE 
instrument showed that bursts were isotropically distributed, strongly 
suggesting either a cosmological or an extended galactic halo distribution, with
essentially zero dipole and quadrupole components. The spectra are definitely 
non-thermal, typically fitted in the MeV range by broken power-laws whose energy per 
decade $\nu\Fnu$ peak is in the range 50-500 KeV \cite{band93}, the power law 
sometimes extending to GeV energies \cite{hur94}. GRB appeared to leave no detectable 
traces at other wavelengths, except in some cases briefly in X-rays 
\cite{stro98,conhu98}.
The gamma-ray durations range from $10^{-3}$ s to about $10^3$ s, with a roughly
bimodal distribution of long bursts of $t_b \simg 2$ s and short bursts of 
$t_b \siml 2$s \cite{kou93}, and substructure sometimes down to milliseconds.
The gamma-ray light curves range from smooth, fast-rise and quasi-exponential decay 
(FREDs), through curves with several peaks, to highly variable curves with  many 
peaks \cite{fm95,kou98}. The pulse distribution is complex \cite{pen96,nor98}, and 
the time histories can provide clues for the geometry of the emitting regions 
\cite{fen96,fen98}. 

Theoretically, it was clear from early on that if GRB are cosmological, enormous 
energies are liberated in a small volume in a very short time, and an 
$e^\pm -\gamma$ fireball must form \cite{pac86,goo86,sp90}, which would expand
relativistically. The main difficulty with this was that a smoothly  expanding
fireball would  convert most of its energy into kinetic energy of accelerated 
baryons (rather than photons), and would produce a quasi-thermal spectrum, while
the typical timescales would not explain events much longer than milliseconds. 
This problem was solved with the introduction of the ``fireball shock model" 
\cite{rm92,mr93a}, based on the realization that shocks are likely to arise,
at the latest when the fireball runs into an external medium, which would 
occur after the fireball is optically thin and would reconvert the kinetic energy 
into nonthermal radiation. The complicated light curves can be understood in terms 
of internal shocks \cite{rm94} in the outflow itself, before it runs into the 
external medium, caused by velocity variations in the outflow from the source, 

The next major breakthrough came in early 1997 when the Italian-Dutch satellite 
Beppo-SAX succeeded in providing accurate X-ray measurements which, after a delay of 
4-6 hours for processing, led to positions \cite{cos97}, 
allowing follow-ups at optical and other wavelengths, e.g. \cite{jvp97}. This paved 
the way for the measurement of redshift distances, the identification of candidate 
host galaxies, and the confirmation that they were indeed at cosmological distances 
\cite{metz97,djo98_0703,kul98}. The detection of other GRB afterglows followed in 
rapid succession, sometimes extending to radio \cite{fra97,fra98} and over timescales 
of many months \cite{jvp98}, and in a number of cases resulted in the identification 
of candidate host galaxies, e.g. \cite{sah97,bloo98_0508,ode98_1214}, etc.  The study 
of afterglows has provided strong confirmation for the generic fireball shock model of 
GRB.  This model in fact led to a correct prediction \cite{mr97a}, in advance of the 
observations, of the quantitative nature of afterglows at wavelengths longer than
$\gamma$-rays, which were in substantial agreement with the data
\cite{vie97a,tav97,wax97a,rei97,wrm97}.

A major issue raised by the measurement of large redshifts, e.g. \cite{kul98,kul99}, 
is that the measured $\gamma$-ray fluences imply a total energy of order
$10^{54}(\Omega_\gamma /4\pi)$ ergs, where $\Delta \Omega_\gamma$ is the solid 
angle into which the gamma-rays are beamed. A beamed jet would clearly alleviate 
the energy requirements, but it is only recently that tentative evidence has been
reported for evidence of a jet \cite{kul99,fru99,cas99}. Whether a jet is present or 
not, such energies are possible \cite{mr97b} in the context of compact mergers
involving neutron star-neutron star (\nsns) or black hole-neutron star
(\bhns) binaries, or in hypernova/collapsar models involving a massive stellar
progenitor \cite{pac98,pop99}. In both cases, one is led to rely on MHD 
extraction of the spin energy of a disrupted torus and/or a central
fast spinning BH, which can power a relativistic fireball resulting in the
observed radiation. 

While it is at present unclear which, if any, of these
progenitors is responsible for GRB, or whether perhaps different progenitors
represent different subclasses of GRB, there is general agreement that they
all would be expected to lead to the generic fireball shock scenario
mentioned above. Much of the current effort is dedicated to understanding
the progenitors more specifically, and trying to determine what effect, if any, 
they have on the observable burst and afterglow.

\section{Black Hole/Debris systems as generic GRB energy sources}
\label{sec:progen}

It has become increasingly apparent in the last few years  that {\it most}
plausible GRB progenitors suggested so far are expected to lead to a 
system with a central BH plus a temporary debris torus around it. Scenarios
leading to this include, e.g. NS-NS or NS-BH mergers, Helium core - black hole 
[He/BH] or white dwarf - black hole [WD-BH] mergers, and the wide category 
labeled as hypernova or collapsars including failed supernova Ib [SNe Ib], 
single or binary Wolf-Rayet [WR] collapse, etc. \cite{pac91,woo93,mr97b,pac98,pop99}, 
and accretion-induced collapse \cite{vs98,pfb99}. An important point is that the 
overall energetics from these various progenitors do not differ by more than about 
one order of magnitude \cite{mrw98b}. Another possibility is massive black holes 
($\sim 10^3 - 10^5 \msun$) in the halos of galaxies.
Some related models involve a compact binary 
or a temporarily rotationally stabilized neutron star, perhaps with a superstrong 
field, e.g. \cite{us94,tho94,vs98,spr99}, which ultimately also should lead to a 
BH plus debris torus.
 
Two large reservoirs of energy are available in these systems: the binding 
energy of the orbiting debris, and the spin energy of the black hole 
\cite{mr97b}. The first can provide up to 42\% of the rest mass energy of 
the disk, for a maximally rotating black hole, while the second can provide 
up to 29\% of the rest mass of the black hole itself. The question is how
to extract this energy. 

One energy extraction mechanisms is the $\nu\bar\nu \to e^+ e^-$ process 
\cite{eic89}, which can tap the thermal energy of the torus produced 
by viscous dissipation. To be efficient, the neutrinos must escape before 
being advected into the hole;
on the other hand, the efficiency of conversion into pairs (which scales with
the square of the neutrino density) is low if the neutrino production is too
gradual. Typical estimates suggest a  fireball of $\siml 10^{51}$ erg
\cite{ruf97,fw98,mcfw99}, except perhaps in the collapsar case where 
\cite{pop99} estimate $10^{52.3}$ ergs for optimum parameters. If the fireball 
is collimated into a solid angle $\Omj$ then of course the apparent 
``isotropized" energy would be larger by a factor $(4\pi/\Omj)$ , but unless 
$\Omj$ is $\siml 10^{-3}-10^{-4}$ this would fail to satisfy the apparent 
isotropized energy of $4\times 10^{54}$ ergs deduced for GRB 990123 
\cite{kul99}.

An alternative, and more efficient mechanism for tapping the energy of the
torus may be through dissipation of magnetic fields generated by the 
differential rotation in the torus \cite{pac91,napapi92,mr97b,ka97}.
Even before the BH forms, a NS-NS merging system might lead to winding up of the
fields and dissipation in the last stages before the merger \cite{mr92,vie97a}.

However, a larger energy source is available in the hole itself, especially
if formed from a coalescing compact binary, since then it is guaranteed to be
rapidly spinning. Being more massive, it could contain more energy than
the torus. The energy extractable in principle through MHD coupling to the
rotation of the hole by the B-Z (Blandford \& Znajek \cite{bz77}) effect could then 
be even larger than that contained in the orbiting debris \cite{mr97b,pac98}.
Collectively, any such MHD outflows have been referred to as Poynting jets.

The various progenitors differ only slightly in the mass of the BH and
that of the debris torus they produce, but they may differ more markedly
in the amount of rotational energy contained in the BH. Strong magnetic
fields, of order $10^{15}$ G, are needed needed to carry away the rotational 
or gravitational energy in a time scale of tens of seconds \cite{us94,tho94},
which may be generated on such timescales by a convective dynamo mechanism,
the conditions for which are satisfied in freshly collapsed neutron stars
or neutron star tori \cite{dt92,klurud98}.
If the magnetic fields do not thread the BH,
then a Poynting outflow can at most carry the gravitational binding energy
of the torus. For a maximally rotating and for a slow-rotating BH this is
\beq
E_{t} = \eps \msun c^2 
\cases{ 0.42 (M_d/\msun)~\hbox{ergs}~,& (fast rot.);\cr
        0.06 (M_d/\msun)~\hbox{ergs}~,& (slow rot.),\cr} ,
\label{eq:edisk}
\enq
where $\eps$ is the efficiency in converting gravitational into MHD jet energy.  
The torus or disk mass in a NS-NS merger is\cite{ruja98} $M_d\sim 
10^{-1}-10^{-2}\msun$ , and for a NS-BH, a He-BH, WD-BH merger or a binary 
WR collapse it may be estimated at \cite{pac98,fw98} $M_d \sim 1\msun$.
In the HeWD-BH merger and WR collapse the mass of the disk is uncertain due 
to lack of calculations on continued accretion from the envelope, so $1\msun$ 
is just a rough estimate. The maximum torus-based MHD energy extraction is
then
\bea
E_{max,t} \sim \cases{ 
 8\times 10^{53} \eps (M_d/\msun)~\hbox{ergs}~,& ; \nonum \cr
 1.2\times 10^{53}\eps (M_d/\msun)~\hbox{ergs}~,& \cr
 0.8\times 10^{53} \eps (M_d/0.1\msun)~\hbox{ergs}~,& . \nonum } 
\label{eq:ediskcases}
\ena 
for the NS-BH, He/WD-BH or collapsar case; the (slow rotating) failed SN Ib 
case; and NS-NS case, respectively.

If the magnetic fields in the torus thread the BH, the spin energy of the BH 
can in principle be extracted via the \cite{bz77} (B-Z) mechanism 
(\cite{mr97b}).  The extractable energy is
\beq
E_{bh} \sim \eps f(a)\Mbh c^2~,
\enq
where $\eps$ is the MHD efficiency factor, $f(a)=1-([1+\sqrt{1-a^2}]/2 )^{1/2}
\leq 0.29$ is the rotational efficiency factor, and $a = Jc/G M^2$ is the rotation 
parameter, which equals 1 for a maximally rotating black hole. The $f(a)$
rotational factor is is small unless $a$ is close to 1, where it rises sharply 
to its maximum value $f(1)=0.29$, so the main requirement is a rapidly 
rotating black hole, $a \simg 0.5$.  For a maximally rotating BH, the 
extractable energy is therefore
\beq
E_{max,bh}\sim
0.29 \eps\Mbh c^2 \sim 5\times 10^{53}\eps (\Mbh/\msun)~\hbox{ergs}.
\enq
Rapid rotation is guaranteed in a  NS-NS merger, since the radius (especially 
for a soft equation of state) is close to that of a black hole and the final 
orbital spin period is close to the required maximal spin rotation period. 
The central BH will have a mass \cite{ruf97,ruja98} of about $2.5 \msun$, 
so the NS-NS system can power a jet of up to $E_{\nsns}\siml 1.3 \times 10^{54} 
\eps (\Mbh/2.5\msun)$ ergs.
A maximal rotation rate may also be possible in a He-BH merger, depending
on what fraction of the He core gets accreted along the rotation axis as
opposed to along the equator \cite{fw98}, and the same should
apply to the binary fast-rotating WR scenario, which probably does not
differ much in its final details from the He-BH merger. For a fast rotating
BH of $2.5-3\msun$ threaded by the magnetic field, the maximal energy
carried out by the jet is then similar or somewhat larger than in the NS-NS case.
The scenarios less likely to produce a fast rotating BH are the NS-BH merger
(where the rotation parameter could be limited to $a \leq M_{ns}/\Mbh$,
unless the BH is already fast-rotating) and the failed SNe Ib (where the
last material to fall in would have maximum angular momentum, but the
material that was initially close to the hole has less angular momentum).
The electromagnetic energy extraction from the BH in these could be limited
by the $f(a)$ factor, but a lower limit would be given by the energy
available from the gravitational energy of the disk, in the second line
of equation (\ref{eq:edisk}).
\begin{figure}[htb]
\centering
\hspace*{-1cm}
\vspace*{-1cm}
\epsfig{figure=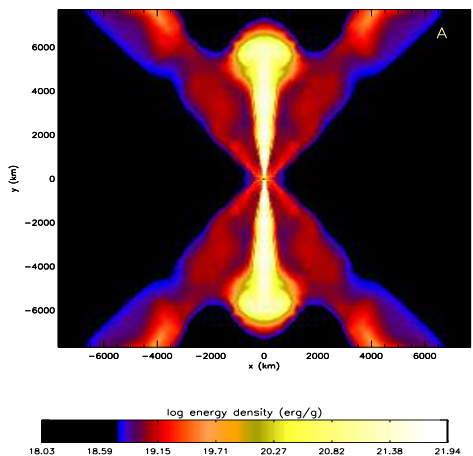, width=6cm, height=6cm}
\caption{Jet formation in a collapsar model leading to black hole from 
collapse of a fast rotating He core \cite{mcfw99}}
   \label{fig:mcfw_jet}
\end{figure}

The total energetics differ thus between the various models at most
by a factor 20 for a Poynting (MHD) jet powered by the torus binding energy, 
whereas for Poynting jets powered by the BH spin energy they differ at most by 
a factor of a few, depending on the
rotation parameter. For instance, allowing for a total efficiency of 50\%, a 
NS-NS merger whose jet is powered by the torus binding energy would require 
a beaming of the $\gamma$-rays by a factor $(4\pi/\Omj)\sim 100$, or beaming 
by a factor $\sim 10$ if the jet is powered by the B-Z mechanism,
to produce the equivalent of an isotropic energy of $4\times 10^{54}$ ergs.  
The beaming requirements of BH-NS and some of the collapsar scenarios are
even less constraining, either when tapping the torus or the BH. Thus, even 
the most extreme energy requirements inferred observationally thus far can be 
plausibly satisfied by scenarios leading to a BH plus torus system.

The major difference between the various models is expected to be in the 
{\it location} where the burst occurs relative to the host galaxy 
(see \S \ref{sec:env}). They are also likely to differ substantially in 
the efficiency of producing a directly observable relativistic outflow, as well 
as in the amount of collimation of the jet they produce. The conditions for the 
efficient escape of a high-$\Gamma$ jet are less propitious if the ``engine" is
surrounded by an extensive envelope. In this case the jet has to ``punch
through" the envelope, and its ability to do so may be crucially dependent
on the level of viscosity achieved in the debris torus (e.g. \cite{mcfw99}),
higher viscosities leading to more powerful jets 
(see Figure \ref{fig:mcfw_jet}). 
The simulations, so far,
are nonrelativistic and one can only infer that high enough viscosities
can lead to jets capable of punching though a massive (several $\msun$)
envelope. This is facilitated, of course, if the envelope is fast-rotating,
as in this case there is a centrifugally induced column density minimum
along the spin axis, which might be small enough to allow punch-through
to occur. If they do, a very tightly collimated beam may arise.
``Cleaner" environments, such as NS-BH or NS-NS merger, or rotational
support loss/accretion induced collapse to BH would have much less
material to be pushed out of the way by a jet, while their energy is,
to order of magnitude, similar to that in massive stellar progenitor cases.
In these cases, on the other hand, there is no natural choke to collimate a 
jet, which might therefore be somewhat wider than in massive progenitor cases.

\section{The Fireball Shock Scenario}
\label{sec:fball}

Irrespective of the details of the progenitor, the resulting fireball
is expected to be initially highly  optically thick. From causality
considerations the initial dimensions must be of order $c t_{var} \siml 10^7$
cm, where $t_{var}$ is the variability timescale,
and the luminosities must be much higher than a solar Eddington limit.
Since most of the spectral energy is observed above 0.5 MeV, the optical
depth against $\gamma\gamma \to e^\pm$ is large, and an $e^\pm,\gamma$
fireball is expected. Due to the highly super-Eddington luminosity, this
fireball must expand. Since in many bursts one observes a large fraction of
the total energy at photon energies $\eps_\gamma \simg 1 GeV$, somehow the
flow must be able to avoid degrading these photons ($\gamma\gamma \to e^\pm$
would lead, in a stationary or slowly expanding flow, to photons just below
0.511 MeV\cite{hb94} ).
In order to avoid this, it seems inescapable that the flow must be expanding
with a very high Lorentz factor, since in this case the relative angle at which
the photons collide is less than $\Gamma^{-1}$ and the threshold for the pair 
production is effectively diminished. The bulk Lorentz factor must be
\beq
\Gamma \simg 10^2 (\eps_{\gamma,\rm 10 GeV} \eps_{t, \rm MeV} )^{1/2}~,
\enq
in order for photons with energy $\eps_\gamma \simg 10$ GeV to escape annihilation 
against target photons of energy $\eps_t \sim 1$ MeV \cite{m95,hb94}.
Thus, simply from observations and general physical considerations, 
a relativistically expanding fireball is expected. 
>From general considerations \cite{mlr93}, one can see that an outflow arising 
from an initial energy $E_o$ imparted to a mass $M_o << E_0/c^2$ within a 
radius $r_l$ will lead to an expansion. Initially the bulk Lorentz $\Gamma 
\propto r$, while comoving temperature drops $\propto r^{-1}$; however,
$\Gamma$ cannot increase beyond $\Gamma_{max} \sim \eta \sim E_o/M_o c^2$, 
which is achieved at a radius $r/r_l \sim \eta$, beyond which the flow 
continues to coast with $\Gamma \sim \eta \sim $ constant \cite{mlr93}.
\beq
\Gamma \sim \cases{ (r/r_l) ~,& for $r/r_l \siml \eta$;\cr
                     \eta   ~,& for $r/r_l \simg \eta$.\cr} .
\enq
However, the observed $\gamma$-ray spectrum
observed is generally a broken power law, i.e., highly nonthermal. The
optically thick $e^\pm \gamma$ fireball cannot, by itself, produce such a
spectrum (it would tend rather to produce a modified blackbody, 
\cite{pac86,goo86}). In addition,
the expansion would lead to a conversion of internal energy into kinetic
energy of expansion, so even after the fireball becomes optically thin,
it would be highly inefficient, most of the energy being in the kinetic
energy of the associated protons, rather than in photons.
 
The most likely way to achieve a nonthermal spectrum in an energetically
efficient manner is if the kinetic energy of the flow is re-converted into
random energy via shocks, after the flow has become optically thin \cite{rm92}. 
This is a plausible scenario, in which two cases can be distinguished. In the 
first case (a) the expanding fireball runs into an external medium (the ISM, or a 
pre-ejected stellar wind\cite{rm92,mr93a,ka94a,sapi95}. The second
possibility (b) is that \cite{rm94,px94}, even before external shocks occur, 
internal shocks develop in the relativistic wind itself, faster portions of the
flow catching up with the slower portions.
This is a completely generic model, which is independent of the specific
nature of the progenitor, as long as it delivers the appropriate amount
of energy ($\simg 10^{52}$ erg) in a small enough region ($\siml 10^7$ cm).
This model has been successful in explaining the major observational
properties of the gamma-ray emission, and is the main paradigm used for
interpreting the GRB observations.
 
External shocks will occur in an impulsive outflow of total energy
$E_o$ in an external medium of average particle density $n_o$ at a radius
and on a timescale
\bea
r_{dec} \sim & 10^{17} E_{53}^{1/3} n_o^{-1/3} \eta_2^{-2/3} ~{\rm cm}~,\nonum \cr
t_{dec} \sim & r_{dec}/(c\Gamma^2) \sim 3\times 10^2  E_{53}^{1/3} n_o^{-1/3}\eta_2^{-8/3} ~{\rm s}~,\cr
\label{eq:rdec}
\ena
where the lab-frame energy of the swept-up external matter ($\Gamma^2 m_p c^2$
per proton) equals the initial energy $E_o$ of the fireball, and $\eta=\Gamma =
10^2\eta_2$ is the final bulk Lorentz factor of the ejecta.
The typical observer-frame dynamic time of the shock (assuming the cooling
time is shorter than this) is $t_{dec} \sim r_{dec}/c \Gamma^2 \sim$ seconds,
for typical parameters, and $t_b \sim t_{dec}$ would be the burst duration (the
impulsive assumption requires that the initial energy input occur in a time
shorter than $t_{dyn}$). Variability on timescales shorter than $t_{dec}$
may occur on the cooling timescale or on the dynamic timescale for
inhomogeneities in the external medium, but generally this is not ideal for
reproducing highly variable profiles\cite{sapi98}.
However, it can reproduce bursts with several peaks\cite{pm98a}
and may therefore be applicable to the class of long, smooth bursts.
 
The same behavior $\Gamma \propto r$ with comoving temperature $\propto 
r^{-1}$, followed by saturation $\Gamma_{max} \sim \eta$ at the same radius 
$r/r_l \sim \eta$ occurs in a wind scenario \cite{pac90}, if one assumes 
that a lab-frame luminosity $L_o$ and mass outflow $\dot M_o$ are injected 
at $r\sim r_l$ and continuously maintained over a time
$t_w$; here $\eta=L_o/ {\dot M_o c^2}$. In such wind model, internal shocks
will occur at a radius and over a timescale \cite{rm94}
\bea
r_{dis} \sim & c t_{var} \eta^2 \sim 3\times 10^{14} t_{var} \eta_2^2 ~
                                                           {\rm cm},\nonum\cr
t_w \gg & t_{var} \sim r_{dis}/(c\eta^2) ~{\rm s},\cr
\label{eq:rdis}
\ena
where shells of different energies $\Delta \eta \sim \eta$ initially separated
by $c t_v $ (where $t_v \leq t_w$ is the timescale of typical variations in
the energy at $r_l$) catch up with each other.
In order for internal shocks to occur above
the wind photosphere $r_{ph} \sim {\dot M} \sigma_T /(4\pi m_p c \Gamma^2)$
$=1.2\times 10^{14} L_{53}\eta_2^{-3}$ cm, but also at radii greater than the
saturation radius (so that most of the energy does not come out in the
photospheric quasi-thermal radiation component) one needs to have
$7.5\times 10^1 L_{51}^{1/5} t_{var}^{-1/5} \siml \eta
3\times 10^2 L_{53}^{1/4} t_{var}^{-1/4}$.
This type of models have the advantage\cite{rm94} that they allow an
arbitrarily complicated light curve, the shortest variation timescale $t_{var} 
\simg 10^{-3}$ s being limited only by the dynamic timescale at $r_l$, where 
the energy input may be expected to vary chaotically.
Such internal shocks have been shown explicitly  to
reproduce (and be required by) some of the more complicated
light curves\cite{sapi98,kps98,pm99int} (see however \cite{dermit98}).

\section{The Simple Standard Afterglow Model} 
\label{sec:staaft}

The dynamics of GRB and their afterglows can be understood in a fairly simple
manner, independently of any uncertainties about the progenitor systems, using 
a generalization of the method used to model supernova remnants. The simplest 
hypothesis is that the afterglow is due to a relativistic expanding blast wave, 
which decelerates as time goes on \cite{mr97a}. 
The complex time structure of some bursts suggests that the central trigger may 
continue for up to 100 seconds, the $\gamma$-rays possibly being due to
internal shocks. However, at much later times all memory of the initial time 
structure would be lost: essentially all that matters is how much energy and 
momentum has been injected; the injection can be regarded as instantaneous in 
the context of the much longer afterglow. 
As pointed in the original fireball shock paper \cite{rm92}, the external shock 
bolometric luminosity builds up and decays as 
\beq
L \propto \cases{ t^2        & rise \nonum \cr
                  t^{-(1+q)} & decay ~.\cr }
\label{eq:Lrise}
\enq
The first line is obtained equating, in the contact discontinuity frame, the kinetic 
flux $L/4\pi r^2 $ to the external ram pressure $\rho_{ext} \Gamma^2$ during the 
initial phase where $\Gamma\sim$ constant, $r\propto t$, while the second 
follows from energy conservation $L\propto E/t$ under adiabatic conditions 
($q$ takes into account radiative effects or bolometric corrections; the flux 
per unit frequency rises in the same way, and decays with $q\geq 1$ in 
equ. (\ref{eq:Lrise})).
At the deceleration radius (\ref{eq:rdec}) the fireball energy and the bulk Lorentz
factor decrease by a factor $\sim 2$ over a timescale $t_{dec}\sim 
r_{dec}/(c\Gamma^2)$, and thereafter the bulk Lorentz factor decreases as a
power law in radius. This is
\beq
\Gamma \propto r^{-g}\propto t^{-g/(1+2g)}~,~r\propto t^{1/(1+2g)},
\label{eq:Gamma}
\enq
with $g=(3,3/2)$ for the radiative (adiabatic) regime, in which 
$\rho r^3 \Gamma \sim$ constant ($\rho r^3 \Gamma^2 \sim$ constant).  
At late times, a similarity solution \cite{bm76} solution with
$g=7/2$ may be reached.

The spectrum of radiation is likely to be due to synchrotron radiation, whose
peak frequency in the observer frame is $\nu_m \propto \Gamma B' \gamma^2$,
and both the comoving field $B'$ and electron Lorentz factor $\gamma$ are
likely to be proportional to $\Gamma$ \cite{mr93a}. This implies that as 
$\Gamma$ decreases, so will $\nu_m$, and the radiation will move to longer
wavelengths. This led \cite{pacro93,ka94b} to early discussions, based on the 
forward blast wave, of the possibility of detecting at late times a radio or 
optical afterglow of the GRB. A more detailed treatment of the fireball dynamics 
indicate that approximately equal amounts of energy are radiated by the forward 
blast wave, moving with $\sim\Gamma$ into the surrounding medium, and by a 
reverse shock propagating with $\Gamma_r -1 \sim 1$ back into the ejecta
\cite{mr93a}. The electrons are therefore shocked to much higher energies in
the forward shock than in the reverse shock, producing a two-step synchrotron
spectrum which during the deceleration time $t_{dec}$ peaks in the optical 
(reverse) and in the $\gamma/X$ (forward) \cite{mr93b,mrp94}. The predicted
fluences in the optical for typical bursts at cosmological distances were
$\sim 10^{-7.5}$ erg cm$^{-2}$ s$^{-1}$, or about a 9th magnitude prompt
optical flash \cite{mr97a} of duration comparable to the $\gamma$-rays, in 
agreement with a recent prompt optical detection in GRB 990123 \cite{ake99}.
Detailed calculations and predictions of the time evolution of such a 
forward and reverse shock afterglow model (\cite{mr97a}) preceded the 
observations of the first afterglow GRB970228 (\cite{cos97,jvp97}), which
was detected in $\gamma$-rays, X-rays and several optical bands, and was
followed up for a number of months.
 
\begin{figure}[htb]
\centering
\hspace*{-1cm}
\vspace*{-1cm}
\epsfig{figure=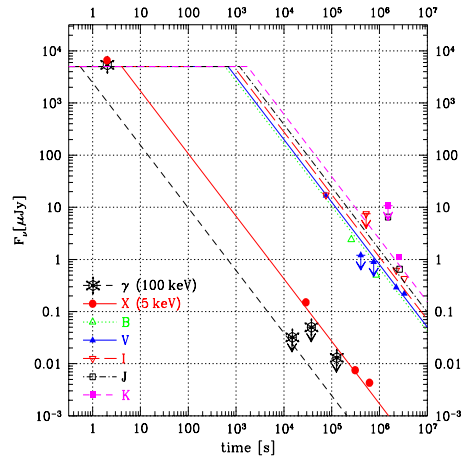, width=6cm, height=6cm}
\caption{GRB 970228 light-curves compared \cite{wrm97} to the blast 
 wave model predictions of \cite{mr97a} }
   \label{fig:0228_lightcurve}
\end{figure}

The simplest spherical afterglow model generally concentrates only on the 
properties of the forward blast wave radiation, for which the flux at a given 
frequency and the synchrotron peak frequency decay at a rate \cite{mr97a,mr99}
\beq
\Fnu\propto  t^{[3-2g(1-2\beta)]/(1+2g)}~~,~~\nu_m\propto t^{-4g/(1+2g)},
\label{eq:Fnu}
\enq
where $g$ is the exponent of $\Gamma$ (equ. [\ref{eq:Gamma}]) and $\beta$ is 
the photon spectral energy slope. The decay rate of the forward shock $\Fnu$ 
in equ.(\ref{eq:Fnu}) is typically slower than that of the reverse shock
\cite{mr97a}, and the reason why the "simplest" model was stripped down to
its forward shock component only is that, for the first two years 1997-1998,
afterglows were followed in more detail only after the several hours needed by 
Beppo-SAX to acquire accurate positions, by which time both reverse external 
shock and internal shock components are expected to have become unobservable. 
This simple standard model has been remarkably successful at explaining
the gross features and light curves of GRB 970228, GRB 970508 (after 2 days; for
early rise, see \S \ref{sec:postaft})
e.g. \cite{wrm97,tav97,wax97a,rei97} (see Figure \ref{fig:0228_lightcurve}).

\begin{figure}[htb]
\centering
\hspace*{-1cm}
\vspace*{-1cm}
\epsfig{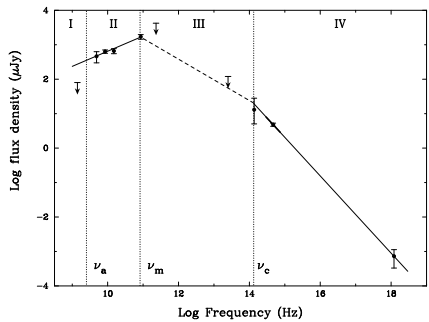}
\caption{Snapshot spectrum of GRB 970508 at $t=12$ days and
 standard afterglow model fit \cite{wiga98}}
   \label{fig:0508_spec}
\end{figure}

This simplest afterglow model produces at any given time a three-segment power 
law spectrum with two breaks. At low frequencies there is a steeply rising
synchrotron self-absorbed spectrum up to a self-absorption break $\nu_a$,
followed by a +1/3 energy index spectrum up to the synchrotron break $\nu_m$
corresponding to the minimum energy $\gamma_m$ of the power-law accelerated
electrons, and then a $-(p-1)/2$ energy spectrum above this break,
for electrons in the adiabatic regime (where $\gamma^{-p}$ is the electron
energy distribution above $\gamma_m$). A fourth segment and a third break is
expected at energies where the electron cooling time becomes short compared
to the expansion time, with a spectral slope $-p/2$ above that. With
this third ``cooling" break $\nu_b$, first calculated in \cite{mrw98} and
more explicitly detailed in \cite{spn98}, one has what has come to be called
the simple ``standard" model of GRB afterglows. One of the predictions of this
model \cite{mr97a} is that the relation between the temporal decay index $\alpha$,
for $g=3/2$ in $\Gamma\propto r^{-g}$, is related to the photon spectral energy 
index $\beta$ through
through
\beq
\Fnu \propto t^\alpha \nu^\beta~~,\hbox{with}~~\alpha=(3/2)\beta~.
\label{eq:alphast}
\enq
This relationship appears to be valid in many (although not all) cases, especially
after the first few days, and is compatible with an electron spectral index $p\sim 
2.2-2.5$ which is typical of shock acceleration, e.g. \cite{wax97a,spn98,wiga98}, 
etc.  As the remnant expands the photon spectrum moves to lower frequencies, and
the flux in a given band decays as a power law in time, whose index can change
as breaks move through it.
For the simple standard model, snapshot overall spectra have been deduced
by extrapolating spectra at different wavebands and times using assumed
simple time dependences \cite{wax97b,wiga98}. These can be used to derive rough
fits for the different physical parameters of the burst and
environment, e.g. the total energy $E$, the magnetic and electron-proton
coupling parameters ${\eps}_B$ and ${\eps}_e$ and the external density $n_o$
(see Figure \ref{fig:0508_spec}).

Since the simple afterglow model has generally proved quite robust, it is worth 
reviewing the assumptions made in it. The following apply both to the ``simple 
standard" model using forward shocks only, and to the original \cite{mr97a} 
version including both forward and reverse shocks:\\
a) A single value of $E_o$ and $\Gamma_o=\eta$ is used,\\
b) the external medium $n_{ext}$ is homogeneous,\\
c) the accelerated electron spectral index $p$, the magnetic field and electron
to proton equipartion ratios $\varep_B$ and $\varep_e$ do not change in time,\\
d) the expansion is relativistic and the dynamics are given by $\Gamma \propto 
r^{-3/2}$ (adiabatic), \\
d) the outflow is spherical (or angle independent inside some jet solid angle 
$\Omj$),\\
e) the observed radiation is characterized by the scaling relations along the 
 line of sight.\\

These assumptions, even if correct over some range, clearly would break down after 
some time.  Estimates for the time needed to reach the non-relativistic expansion
regime are typically $\siml$ month(s) (\cite{vie97a}), or less if there is an initial 
radiative regime $\Gamma\propto r^{-3}$.  However, even when electron radiative times 
are shorter than the expansion time, it is unclear whether a regime $\Gamma\propto 
r^{-3}$ should occur, since it would require strong electron-proton coupling 
\cite{mrw98}. As far as sphericity, the standard model can be straightforwardly 
generalized to the case where the energy is assumed to be channeled initially into a
solid angle $\Omj < 4\pi$ \cite{mlr93}. In this case \cite{rho97,rho99} a change
occurs after $\Gamma$ drops below $\Omj^{-1/2}$, after which the side of the jet
becomes observable, and soon thereafter one expects a faster decay of $\Gamma$ 
if the jet starts to expands sideways, leading to a decrease in the brightness. 
A calculation based on the sideways expansion, using the usual scaling laws for a 
single central line of sight \cite{rho99} leads then to a steepening of
the light curve. Until recently, no evidence for a steepening could be found
in afterglows over several months. E.g., in GRB 971214 \cite{ram98}, a
snapshot standard model fit and the lack of a break in the late light curve 
could be, in principle, interpreted as evidence for lack of a jet, leading
to an (isotropic) energy estimate of $10^{53.5}$ ergs. While such large energy
outputs are possible in {\it either} NS-NS, NS-BH mergers \cite{mr97b} or in 
hypernova/collapsar models \cite{pac98,pop99} using MHD extraction of the 
spin energy of a disrupted torus and/or a central fast spinning BH, it is worth 
stressing that what these snapshot fits constrain is only the {\it energy per 
solid angle} \cite{mrw98b}. Also, the expectation of a break after some weeks 
or months (e.g., due to $\Gamma$ dropping either below a few, or below 
$\Omega_j^{-1/2}$) is based upon the simple impulsive (angle-independent delta 
or top-hat function) energy input approximation. The latter is useful, but 
departures from it would be natural, and certainly not surprising. In fact,
as discussed below, tentative evidence for beaming in one obejct has recently
been reported \cite{kul99,fru99,cas99}, but it is difficulty to ascertain, and 
could be masked by a number of commonly expected effects.

\section{``Post-standard" Afterglow Models}
\label{sec:postaft}
 
\begin{figure}
\vskip -0.3in
\hskip .1in   \resizebox{\hsize}{!}{\includegraphics{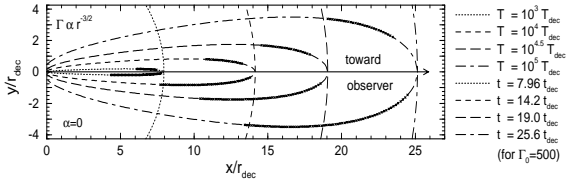}}
\vskip -1.2in
   \caption{Ring-like equal-arrival time $T$ surfaces of an afterglow, based
   on \cite{pmring98}} 
   \label{fig:ring} 
\end{figure}

\begin{figure}[htb]
\centering
\hspace*{-1cm}
\vspace*{-1cm}
\epsfig{figure=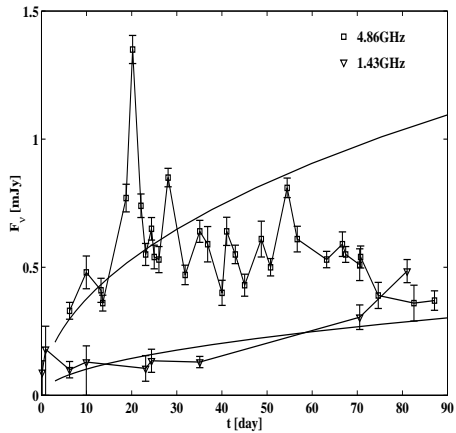, width=6cm, height=6cm}
\caption{Radio afterglow light-curves of GRB970508 at 4.86GHz
and 1.43GHz, compared with the predictions of \cite{wkf98}.}
   \label{fig:scint}
\end{figure}

In a realistic situation, one could expect any of several fairly natural
departures from the simple standard model to occur. The first one is that
the emitting region seen by the observer resembles a ring 
\cite{wax97b,pm98b,sari98} (see Figure \ref{fig:ring}). This effect may,
in fact, be important in giving rise to the radio scintillation
pattern seen in several afterglows, since this requires the emitting source 
to be of small dimensions, which is aided if the emission is ring-like,
e.g. in the example of GRB 970508 \cite{wkf98} (Figure \ref{fig:scint}).

One expects afterglows to show a diversity in their decay rates, not only due to
different $\beta$ but also from the possibility of a non-standard relation between 
the temporal decay index $\alpha$ and the spectral energy index $\beta$, different 
from equ. (\ref{eq:alphast}). The most obvious departure from the simplest
standard model occurs if the external medium is inhomogeneous: for instance, for 
$n_{ext} \propto r^{-d}$, the energy conservation condition is $\Gamma^2 r^{3-d} 
\sim$ constant, which changes significantly the temporal decay rates \cite{mrw98}. 
Such a power law dependence is expected if the external medium is a wind, say from 
an evolved progenitor star as implied in the hypernova scenario (such winds are 
generally used to fit supernova remnant models). Another obvious non-standard effect,
which it is reasonable to expect, is departures from a simple impulsive injection 
approximation (i.e. from a delta or top hat function with a single value for 
$E_o$ and $\Gamma_o$). An example is if the mass and energy injected during the 
burst duration $t_w$ (say tens of seconds) obeys $M(>\Gamma) \propto
\Gamma^{-s}$, $E(>\Gamma)\propto \Gamma^{1-s}$, i.e. more energy emitted with 
lower Lorentz factors at later times (but still shorter than the gamma-ray pulse 
duration). This would drastically change the temporal decay rate and extend the 
afterglow lifetime in the relativistic regime, providing a late ``energy refreshment" 
to the blast wave on time scales comparable to the afterglow time scale 
\cite{rm98}. These two cases lead to a decay rate 
\beq
\Gamma \propto r^{-g} \propto \cases{
  r^{-(3-d)/2} & ~; $n_{ext}\propto r^{-d}$;\cr
  r^{-3/(2+s)} & ~; $E(>\Gamma)\propto \Gamma^{1-s}$.\cr }
\label{eq:Gammanonst}
\enq
Expressions for the temporal decay index $\alpha (\beta,s,d)$ in $\Fnu\propto 
t^\alpha$ are given by \cite{mrw98,rm98}, which now depend also on $s$ and/or $d$ 
(and not just on $\beta$ as in the simple standard relation of equ.(\ref{eq:alphast}).
The result is that the decay can be flatter (or steeper, depending on $s$ and $d$) 
than the simple standard $\alpha= (3/2)\beta$.
A third non-standard effect, which is entirely natural, occurs when the energy
and/or the bulk Lorentz factor injected are some function of the angle. A simple case
is $E_o\propto \theta^{-j}$, $\Gamma_o\propto \theta^{-k}$ within a range of angles; 
this leads to the outflow at different angles shocking at different radii and its 
radiation arriving at the observed at different delayed times, and it has a marked 
effect on the time dependence of the afterglow \cite{mrw98}, with $\alpha=\alpha
(\beta,j,k)$ flatter or steeper than the standard value, depending on $j,k$. 
Thus in general, a temporal decay index which is a function of more than one 
parameter
\beq
\Fnu\propto t^\alpha\nu^\beta~~,\hbox{with}~~\alpha=\alpha (\beta,d,s,j,k,\cdots )~,
\label{eq:alphanonst}
\enq
is not surprising; what is more remarkable is that, in many cases, the simple
standard relation (\ref{eq:alphast}) is sufficient to describe the gross overall 
behavior at late times.

\begin{figure}
\vskip .2 in
\hskip -.5in   \resizebox{\hsize}{!}{\includegraphics{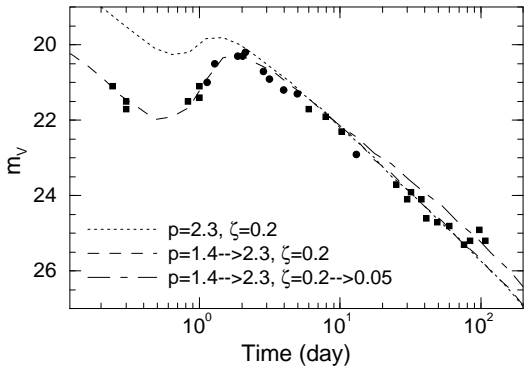}}
   \caption{Optical light-curve of GRB 970508, fitted with a non-uniform
  injection model (a similar fit can be obtained with an off-axis jet plus
  a weaker isotropic component) \cite{pmr98}}
   \label{fig:injfit0508}
\end{figure}

Strong evidence for departures from the simple standard model is provided by,
e.g., sharp rises or humps in the light curves followed by a renewed decay,
as in GRB 970508 (\cite{ped98,pir98a}). Detailed time-dependent model fits
\cite{pmr98} to the X-ray, optical and radio light curves of GRB 970228 and 
GRB 970508 show that, in order to explain the humps, a {\it non-uniform} injection 
(Figure \ref{fig:injfit0508}) or an {\it anisotropic} outflow 
is required. 
 
These fits indicate that
the shock physics may be a function of the shock strength (e.g. the electron
index $p$, injection fraction $\zeta$ and/or $\epsilon_b,~\epsilon_e$ change
in time), and also indicate that dust absorption is needed to simultaneously
fit the X-ray and optical fluxes. The effects of beaming (outflow within a
limited range of solid angles) can be significant \cite{pmjet99}, but are coupled 
with other effects, and a careful analysis is needed to disentangle them.

One consequence of ``post-standard" decay laws (e.g. from density inhomogeneities,
non-uniform injection or anisotropic outflow) is that the transition to a 
steeper jet regime $\Gamma < \theta_j^{-1} \sim$ few can occur as late as six 
months to a year after the outburst, depending on details of the energy input.  
This transition is made more difficult to detect by the fact that, as numerical
integration over angles of the ring-like emission \cite{pmring98} show, the
transition is very gradual and the effects of sideways expansion effects are not 
so drastic as inferred \cite{rho99} from the scaling laws along the central line of 
sight. This is because even though the flux from the head-on part of the remnant
decreases faster, this is more than compensated by the increased emission
measure from sweeping up external matter over a larger angle, and by the
fact that the extra radiation, arising at larger angles, arrives later and
re-fills the steeper light curve. The inference (e.g. \cite{ram98,rho99}) that
GRB 970508 and a few other bursts were isotropic due to the lack of an observable
break is predicated entirely on the validity of the {\it simplest standard} fireball 
assumption. Since these assumptions are drastic simplifications, and physically
plausible generalizations lead to different conclusions, one can interpret the 
results of \cite{ram98,rho99} as arguments indicating that {\it post-standard} 
features are, in fact, necessary in some objects.

\section{Prompt multi-wavelength flashes, reverse shocks and jets}
\label{sec:promptjet}

Prompt optical, X-ray and GeV flashes from reverse and forward shocks, as well as 
from internal shocks, have been calculated in theoretical fireball shock models 
for a number of years \cite{mr93b,mrp94,pm96,mr97a,sp99a}, as have been jets (e.g. 
\cite{mr92,mlr93,mrp94}, and in more detail \cite{rho97,pmr98,pmjet99,rho99}).
Thus, while in recent years they were not explicitly part of the ``simple standard" 
model, they are not strictly ``post"-standard either, since they generally use
the ``standard" assumptions, and they have a long history.
However, observational evidence for these effects were largely lacking, until
the detection of a prompt (within 22 s) optical flash from GRB 990123 with
ROTSE by \cite{ake99}, together with X-ray, optical and radio follow-ups 
cite{kul99,gal99,fru99,and99,cas99,hjo99}. GRB 990123 is so far unique not only
for its prompt optical detection, but also by the fact that if it were emitting
isotropically, based on its redshift $z=1.6$ \cite{kul99,and99} its energy would 
be the largest of any GRB so far, $4\times 10^{54}$ ergs. It is, however, also
the first (tentative) case in which there is evidence for jet-like emission 
\cite{kul99,fru99,cas99}. An additional, uncommon feature is that a radio afterglow 
appeared after only one day, only to disappear the next \cite{gal99,kul99}.

The prompt optical light curve of GRB 990123 decays initially as $\propto t^{-2.5}$ 
to $\propto t^{-1.6}$ \cite{ake99}, much steeper than the typical $\propto 
t^{-1.1}$ of previous optical afterglows detected after several hours.
However, after about 10 minutes its decay rate moderates, and appears to
join smoothly onto a slower decay rate $\propto t^{-1.1}$ measured with 
large telescopes \cite{gal99,kul99,fru99,cas99} after hours and days. The prompt
optical flash peaked at 9-th magnitude after 55 s \cite{ake99}, and in fact a
9-th magnitude prompt flash with a steeper decay rate had been predicted more than 
two years ago \cite{mr97a}, from the synchrotron radiation of the reverse shock
in GRB afterglows at cosmological redshifts (see also {\cite{sp99a}). An optical 
flash contemporaneous with the $\gamma$-ray burst, coming from the reverse shock 
and with fluence corresponding to that magnitude, had also been predicted earlier 
\cite{mr93b,mrp94}. An origin of the optical prompt flash in internal shocks 
\cite{mr97a,mr99} cannot be ruled out yet, but is less likely since the optical 
light curve and the $\gamma$-rays appear not to correlate well \cite{sp99b,gal99} 
(but the early optical light curve has only three points). The subsequent slower 
decay agrees with the predictions for the forward component of the external shock 
\cite{mr97a,sp99b,mr99}.

The evidence for a jet is possibly the most exciting, although must still
be considered tentative. It is based on an apparent steepening of the light curve 
after about three days \cite{kul99,fru99,cas99}. This
is harder to establish than the decay of the two previous earlier portions of
the light curve, since by this time the flux has decreased to a level where
the detector noise and the light of the host galaxy become important. However,
after correcting for this, the r-band data appears to steepen significantly. 
(In the K-band, where the noise level is higher, a steepening is not obvious,
but the issue should be settled with further Space Telescope observations).
If real, this steepening is probably due to the transition between early
relativistic expansion, when the light-cone is narrower than the jet opening,
and the late expansion, when the light-cone has become wider than the jet,
leading to a drop in the effective flux \cite{rho97,kul99,mr99,rho99}. A rough 
estimate leads to a jet opening angle of 3-5 degrees, which would reduce the total 
energy requirements to about $4\times 10^{52}$ ergs. This is about two order of 
magnitude less than the binding energy of a few solar rest masses, which, even 
allowing for substantial inefficiencies, is compatible with currently favored 
scenarios (e.g. \cite{pop99,mcfw99}) based on a stellar collapse or a compact 
binary merger.

\section{Location and Environmental Effects} 
\label{sec:env}
 
The location of the afterglow relative to the host galaxy center can
provide clues both for the nature of the progenitor and for the external
density encountered by the fireball. A hypernova model would be expected
to occur inside a galaxy in a high density environment $n_o > 10^3-10^5$ cm$^{-3}$.
Most of the detected and well identified afterglows are inside the projected image 
of the host galaxy \cite{bloo98rome}, and some also show evidence for a dense 
medium at least in front of the afterglow (\cite{ow98}). 
For a number of bursts there are constraints from the 
lack of a detectable, even faint, host galaxy \cite{sch98}, but at least for
Beppo-SAX bursts (which is sensitive only to long bursts $t_b \simg 20$ s) the
success rate in finding candidate hosts is high.

In NS-NS mergers one would expect a BH plus debris torus system and
roughly the same total energy as in a hypernova model, but the mean distance
traveled from birth is of order several Kpc \cite{bsp99}, leading to a burst 
presumably in a less dense environment. The fits of \cite{wiga98} to the 
observational data on GRB 970508 and GRB 971214 in fact suggest external densities 
in the range of $n_o=$ 0.04--0.4 cm$^{-3}$, which would be more typical of a 
tenuous interstellar medium. These could be within the volume of the galaxy, 
but for NS-NS on average one would expect as many GRB inside as outside. This is 
based on an estimate of the mean NS-NS merger time of $10^8$ years; other estimated
merger times (e.g. $10^7$ years, \cite{vdh92}) would give a burst much closer
to the birth site. BH-NS mergers would also occur in timescales $\siml 10^7$
years, and would be expected to give bursts well inside the host galaxy 
(\cite{bsp99}; see however \cite{fw98}). In at least one ``snapshot" standard
afterglow spectral fit for GRB 980329 \cite{reila98} the deduced external
density is $n_o\sim 10^3$ cm$^{-3}$. In some of the other detected afterglows 
there is other evidence for a relatively dense gaseous environments, as 
suggested, e.g. by evidence for dust \cite{rei98} in GRB970508,
the absence of an optical afterglow and presence of strong soft X-ray
absorption \cite{gro97,mur97} in GRB 970828, the lack an an optical
afterglow in the (radio-detected) afterglow (\cite{tay97}) of GRB980329, and
spectral fits to the low energy portion of the X-ray afterglow of several
bursts \cite{ow98}. The latter observations may be suggestive of hypernova
models \cite{pac98,fw98}, involving the collapse of a massive star or its
merger with a compact companion.  

One important caveat is that all afterglows found so far are based on Beppo-SAX 
positions, which is sensitive only to long bursts $t_b \simg 20$ s \cite{hur98}.
This is significant, since it appears
likely that NS-NS mergers lead \cite{mcfw99} to short bursts with $t_b \siml 10$ s.
To make sure that a population of short GRB afterglows is not being 
missed will probably need to await results from HETE \cite{hetepage} and from the 
planned Swift \cite{swiftpage} mission, which is designed to accurately locate 
300 GRB/yr.

An interesting case is the apparent coincidence of GRB 980425 with the 
unusual SN Ib/Ic 1998bw \cite{gal98_SN}, which may represent a new class of SN
\cite{iwa98,bloomSN98}. If true, this could imply that some or perhaps
all GRB could be associated with SN Ib/Ic \cite{wawe98}, differring only in
their viewing angles relative to a very narrow jet. Alternatively,
the GRB could be (e.g. \cite{wes98}) a new subclass of GRB with 
lower energy $E_\gamma \sim 10^{48} (\Omj /4\pi )$ erg, only rarely observable,
while the great majority of the observed GRB would have the energies $E_\gamma 
\sim 10^{54}(\Omj/4\pi)$ ergs as inferred from high redshift observations.
The difficulties are that it would require extreme collimations 
by factors $10^{-3}-10^{-4}$, and the statistical association is so far not 
significant \cite{kip98}.

The environment in which a GRB occurs should also influence the nature of the 
afterglows in other ways. The blast wave and reverse shock that give rise to the
X-rays, optical, etc occur over timescales proportional to $t_{dec} \propto 
n_{ext}^{-1/3}$ (equ.[\ref{eq:rdec}]) which is longer in lower density environments,
so for the same energy the flux is lower, roughly $\Fnu \propto E_o n_{ext}^{1/2}$,
contributing also to make afterglows in the intergalactic medium harder to detect.
However, in addition to affecting broad-band fluxes, one may also expect specific
spectral signatures from the external medium imprinted in the X-ray and optical
continuum, such as atomic edges and lines \cite{bkt97,pl98,mr98b}. These may be
used both to diagnose the chemical abundances and the ionization state (or local
separation from the burst), as well as serving as potential alternative redshift
indicators. (In addition, the outflowing ejecta itself may also contribute 
blue-shifted edge and line features, especially if metal-rich blobs or filaments are
entrained in the flow from the disrupted progenitor debris \cite{mr98a}, which
could serve as diagnostic for the progenitor composition and outflow Lorentz factor).
To distinguish between progenitors (\S \ref{sec:progen}), an interesting prediction 
(\cite{mr98b}; see also \cite{ghi98,bot98}) is that the presence of a measurable 
Fe  K-$\alpha$ X-ray {\it emission} line could be a diagnostic of a hypernova, 
since in this case one may expect a massive envelope at a radius comparable to a 
light-day where $\tau_T \siml 1$, capable of reprocessing the X-ray continuum by 
recombination and fluorescence.  Two groups \cite{piro98b,yosh98} have in fact
recently reported the possible detection of Fe emission lines in GRB 970508 and 
GRB 970828.

\section{ Conclusions }
\label{sec:concl}
 
The fireball shock model of gamma-ray bursts has proved quite robust in 
providing a consistent overall interpretation of the major features of these 
objects at various frequencies and over timescales ranging from the short
initial burst to afterglows extending over many months. The standard internal
shock scenario is able to reproduce the properties of the $\gamma$-ray light
curves, while external shocks involving a forward blast wave and a reverse shock
are successful in reproducing the afterglows observed in X-rays, optical and radio.
The ``simple standard model" of afterglows, involving four spectral slopes and 
three breaks is quite useful in understanding the `snapshot' multiwavelength 
spectra of most afterglows. However, the effects associated with a jet-like
outflow and the possible differential beaming at various energies requires
further investigations, both theoretical and observational. Caution is required 
in interpreting the observations on the basis of the simple standard model. For 
instance, more detailed numerical models, as opposed to the more common analytical
scaling law models, show that the contributions of radiation from different angles
and the gradual transition between different dynamical and radiative regimes 
lead to a considerable rounding-off of the spectral shoulders and light-curve
slope changes, so that breaks cannot be easily located unless the spectral 
sampling is dense and continuous, both in frequency and in time.
Some of the observed light curves with humps, e.g. in GRB 970508, require
`post-standard' model features (i.e. beyond those assumed in the standard
model), such as either non-uniform injection episodes or anisotropic outflows.
Time-dependent multiwavelength fits \cite{pmr98} of some 
bursts also indicate that the parameters characterizing the shock physics change 
with time. Even without humps or slope changes, a non-standard relation between the
spectral and temporal decay slope is observed in several objects, e.g. in
GRB 990123 \cite{kul99}. These are, in our view \cite{mr99}, a strong indication 
for ''post-standard" effects in such bursts.
 
Much progress has been made in understanding how gamma-rays can arise in
fireballs produced by brief events depositing a large amount of energy in a
small volume, and in deriving the generic properties of the long wavelength
afterglows that follow from this.  There still remain a number of mysteries,
especially concerning the identity of their progenitors, the nature of the 
triggering mechanism, the transport of the energy, the time scales involved,
and the nature and effects of beaming. However, even if we do not yet 
understand the details of the gamma-ray burst central engine, it is clear that
these phenomena are among the most powerful transients in the Universe, and they 
could serve as powerful beacons for probing the high redshift ($z > 5$)
universe. The modeling of the burst mechanism itself, as well as the resulting
outflows and radiation, will continue to be a formidable challenge to theorists 
and to computational techniques. However, the theoretical understanding appears
to be converging, and with dedicated new and planned observational missions under 
way, the prospects for significant progress are realistic.
 
I am grateful to Martin Rees for stimulating collaborations, as well as to Alin
Panaitescu, Hara Papathanassiou and Ralph Wijers.

\end{document}